\documentclass[12pt]{article}

\usepackage{graphicx}
\usepackage{amsmath}
\usepackage{amssymb}
\allowdisplaybreaks

\begin{document}

\baselineskip=17.5pt plus 0.2pt minus 0.1pt

\renewcommand{\theequation}{\arabic{equation}}
\renewcommand{\thefootnote}{\fnsymbol{footnote}}
\makeatletter
\def\CR{\nonumber \\}
\def\pt{\partial}
\def\be{\begin{equation}}
\def\ee{\end{equation}}
\def\bea{\begin{eqnarray}}
\def\eea{\end{eqnarray}}
\def\bead{\be\begin{aligned}}
\def\eead{\end{aligned}\ee}
\def\eq#1{(\ref{#1})}
\def\la{\langle}
\def\ra{\rangle}
\def\hyp{\hbox{-}}


\begin{titlepage}
\title{\hfill\parbox{4cm}{ \normalsize YITP-10-37}\\
\vspace{1cm} A renormalization procedure for tensor models \\
and  scalar-tensor theories of gravity}
\author{
Naoki {\sc Sasakura}\thanks{\tt sasakura@yukawa.kyoto-u.ac.jp}
\\[15pt]
{\it Yukawa Institute for Theoretical Physics, Kyoto University,}\\
{\it Kyoto 606-8502, Japan}}
\date{}
\maketitle
\thispagestyle{empty}
\begin{abstract}
\normalsize
Tensor models are more-index generalizations of the so-called matrix models, 
and provide models of quantum gravity with the idea that spaces and general relativity are emergent phenomena.
In this paper, a renormalization procedure for the tensor models whose dynamical variable is a totally symmetric 
real three-tensor is discussed. It is proven that configurations with certain Gaussian forms are the attractors 
of the three-tensor under the renormalization procedure. Since these Gaussian configurations
are parameterized by a scalar and a symmetric two-tensor, it is argued that, in general situations,
the infrared dynamics of the tensor models 
should be described by scalar-tensor theories of gravity.

\end{abstract}
\end{titlepage}

\section{Introduction}
\label{sec:intro}
In the scenario of emergent spacetime \cite{ref:emergent1,ref:emergent2,ref:emergent3,ref:emergent4,ref:emergent5,ref:emergent6,Ambjorn:2010rx},
the notions of spacetime and distance are not fundamental, but they are just infrared phenomena which appear only after 
underlying dynamics of fundamental degrees of freedom are averaged over.
Therefore, to compare the infrared dynamics of such a scenario with the conventional spacetime formulation of physics such as the general
relativity, 
it is necessary to give an appropriate definition of distance and derive the metric tensor field on an emergent
spacetime. An explicit example of such a definition is given in the noncommutative geometry, where 
the notion of distance is defined through a Laplacian or a Dirac operator on a noncommutative spacetime \cite{ref:connes}. 
Another example is given in the simplicial quantum gravity, where such operational definitions of distance have been used for 
comparisons with the general relativity \cite{Ambjorn:2010rx}.

Because of the infrared character of the notion of distance, different operational definitions of distance are not generally identical 
in short distances but converge in long distances.
This implies that a metric tensor field on an emergent spacetime could be derived more directly from 
application of a renormalization procedure to underlying small distance dynamics of fundamental degrees of 
freedom.\footnote{In fact, in the context of the AdS/CFT correspondence in string theory, 
coarse-graining and renormalization are argued to be essentially important in relating boundary field theories 
to bulk classical geometries \cite{ref:emergent3}.}
In this paper, I will discuss a renormalization procedure for the tensor models, which
are models of emergent space, and see how a metric tensor field emerges.

The tensor models 
were originally proposed to describe the simplicial quantum gravity in more than two
dimensions \cite{Ambjorn:1990ge,Sasakura:1990fs,Godfrey:1990dt}. Although the
subsequent developments \cite{Boulatov:1992vp,Ooguri:1992eb,DePietri:1999bx,DePietri:2000ii,Gurau:2009tw,Gurau:2009tz} 
have demonstrated some interesting connections to the loop and spin foam quantum gravities, 
the tensor models seem to have not yet become a fully successful approach
under the original interpretation of the tensor models based on
the diagrammatic relations between the tensor models and simplicial manifolds
in more than two dimensions.
In view of this status, the author has proposed another interpretation of the tensor models 
as theory describing dynamics of fuzzy spaces \cite{Sasakura:2005js}.
This reinterpretation has the following advantages compared with the original one. 
Firstly, a three-tensor is enough to describe all the dimensional spaces. Secondly,  
the semiclassical analysis of 
the tensor models can be considered to have the physical meaning that
a classical solution and small fluctuations around it correspond to a background space 
and field theories, respectively.
Lastly, the diffeomorphism invariance of the general relativity can naturally be incorporated into the symmetry of the tensor 
models. 
Indeed, the tensor models which are fine-tuned to possess classical solutions with certain Gaussian forms 
have numerically been analyzed, 
and the agreement between the tensor models and the general relativity\footnote{
Here the terminology ``general relativity" is not appropriate in its strict sense.
I use it to mean a theory which has a metric tensor field with Euclidean signature
as its dynamical variable, and has the diffeomorphism symmetry. Moreover, its action may not be the Einstein-Hilbert action, $R$, 
but may be $R^2$, or other invariants.} 
 has been obtained in various 
dimensions \cite{Sasakura:2007sv,Sasakura:2007ud,Sasakura:2008pe,Sasakura:2009hs,Sasakura:2009dk}.

The purpose of this paper is to give a first step to generalize the above results
of the fine-tuned tensor models to more general tensor models.
It will be proven that the Gaussian configurations are the attractors of the three-tensor under a renormalization
procedure. 
Since these Gaussian configurations can be parameterized by a scalar and a symmetric two-tensor, 
it will be argued that, in general situations,  the infrared dynamics of the tensor models will be described in
terms of scalar-tensor theories of gravity.

Coupled field theories of the metric tensor and a scalar
have been considered for various physical reasons \cite{Brans:1996tp}.
For example, motivated by Mach's idea on inertial induction,
the Brans-Dicke theory \cite{Brans:1961sx} has a scalar field which plays as a variable gravitational ``constant".
In string theory, the dilaton field is necessarily introduced to keep the worldsheet conformal invariance 
after quantization \cite{Green:1987sp}. Scalar fields are actively discussed to model the dark energy and 
the dark matters, which have clearly been shown to exist in the universe
by recent astrophysical observations \cite{Tsujikawa:2010sc}. 
Therefore it will be highly interesting, if it is concluded that 
the tensor models necessarily generate a scalar field  as well as the metric tensor field in their infrared dynamics.

This paper is organized as follows. In Section \ref{sec:tensor}, the tensor models in this paper 
and the Gaussian configurations are introduced.
In Section \ref{sec:renormalization}, a renormalization procedure for the tensor models is introduced.
In Section \ref{sec:proof}, it is proven that the Gaussian configurations are the attractors of the three-tensor
under the 
renormalization procedure.
In Section \ref{sec:example}, two numerical examples are given for showing how configurations converge to Gaussian configurations by iterations of the renormalization procedure.
In Section \ref{sec:scalartensor}, it is argued that the infrared dynamics of the tensor models should be described in terms of
scalar-tensor theories of gravity. 
The last section is devoted to summary, discussions and future directions.
In Appendix \ref{app:details}, some details of the proof in Section \ref{sec:proof} are provided.

\section{Tensor models and Gaussian configurations}
\label{sec:tensor}
The general idea of regarding the tensor models as theory describing dynamics of fuzzy spaces is based on the observation that
a fuzzy space is characterized by an algebra of functions on it, 
$f_a\, f_b=C_{ab}{}^c f_c$ \cite{Sasakura:2005js}.\footnote{The algebra considered is generally nonassociative. Otherwise,
an algebra cannot encode geometry of a fuzzy space. For example, 
geometry is encoded in a Laplacian (or a Dirac operator) 
on a noncommutative fuzzy space \cite{ref:connes}, but not in a noncommutative algebra itself.}
By considering the three-tensor $C$ to be a dynamical variable, one can obtain the tensor models as theory of dynamics of fuzzy spaces.
Depending on the detailed choices of a dynamical variable and a symmetry, there exist various kinds of tensor models. 

In this paper, the dynamical variable of the tensor models is a totally symmetric real three-tensor $C_{abc}\ (a,b,c=1,2,\ldots,N)$,
which satisfies
\bead
\label{eq:defc}
&C_{abc} \in \mathbb{R}, \\
&C_{abc}=C_{bca}=C_{cab}=C_{bac}=C_{acb}=C_{cba}. \\
\eead
The tensor models are assumed to be invariant under the orthogonal transformation,
\be
\label{eq:onsym}
C'_{abc}=M_a{}^{a'} M_b{}^{b'} M_c{}^{c'} C_{a'b'c'},\ \ M\in O(N).
\ee
Then an example of an invariant action can be given by
\be
\label{eq:action}
S=-\frac{g_0}{2}\, C_{abc}C_{abc}+\frac{g_1}{4}\, C_{abc}C_{dbc}C_{def}C_{aef}-\frac{g_2}{4}\, C_{abc}C_{ade}C_{bdf}C_{cef},
\ee
where the repeated indices are assumed to be summed 
over\footnote{This will be assumed in the rest of this paper, unless otherwise stated.}, and $g_i$ are real coupling constants.
This is one of the simplest actions, which is known to have a number of classical solutions corresponding to various dimensional
fuzzy spheres and tori \cite{Sasakura:2006pq}. In Section \ref{sec:example}, I will use its classical solutions as examples for applying 
the renormalization procedure introduced in Section \ref{sec:renormalization}. 

Next, let me introduce the Gaussian configurations \cite{Sasakura:2007sv,Sasakura:2007ud,Sasakura:2008pe}. 
In concrete applications, the indices of the three-tensor 
are taken to be discrete and finite, but, as ideal cases, let me assume the indices to be continuous and infinite 
in the following discussions.
In the coordinate basis, the Gaussian configurations are defined to take the form,
\be
\label{eq:cx}
C^g_{x_1\, x_2\, x_3}=B \exp \left[
- \beta \left((x_1-x_2)^2+(x_2-x_3)^2+(x_3-x_1)^2 \right)\right], 
\ee 
where $x_i$ are $D$-dimensional coordinates $x_i\equiv (x_i^1,x_i^2,\ldots,x_i^D)$,
$B$ and $\beta$ are positive real constants, and $x^2\equiv g_{\mu\nu} x^\mu x^\nu$. Here 
$g_{\mu\nu}$ is a constant real symmetric two-tensor and is assumed to be positive-definite
to demand exponential damping of $C^g$ at large mutual distances among $x_i$. 
The parameter $\beta$ is redundant in the sense that it can be absorbed into the rescaling of 
$g_{\mu\nu}$, 
but it will be kept for the later convenience to characterize the order of the fuzziness, 
$1/\sqrt{\beta}$,
while $g_{\mu\nu}\sim O(1)$. 

Because of the obvious translational invariance of \eq{eq:cx}, it is more convenient to express \eq{eq:cx}
in the momentum basis\footnote{In the momentum basis, the contractions of indices must be taken 
for pairs of $p$ and $-p$.}. 
By applying Fourier transformation to the indices $x_i$ of \eq{eq:cx}, one obtains
\be
\label{eq:cp}
C^g_{p_1\, p_2 \, p_3}= A \exp \left[-\alpha \left( (p_1)^2+(p_2)^2+(p_3)^2 \right)\right] \delta^D(p_1+p_2+p_3),
\ee
where $\alpha=1/(12 \beta)$, $A$ is a positive constant proportional to $B$, 
$p_i\equiv (p_i^1,p_i^2,\ldots,p^D_i)$, 
and $p^2\equiv g^{\mu\nu} p_\mu p_\nu$ with $g^{\mu\nu}$ being the inverse of the matrix $g_{\mu\nu}$.

The algebra of functions on a usual continuous space is given by $f_{x_1} f_{x_2}=\delta^D (x_1-x_2)\, f_{x_1}$.
This corresponds to $C_{x_1\,x_2\,x_3}=\delta^D(x_1-x_2)\delta^D(x_2-x_3)$ by identifying $C_{abc}$ 
with the structure constant of the algebra, $f_a f_b=C_{abc} f_c$. 
In \eq{eq:cx}, the delta functions are smoothened by Gaussian functions. In addition,
\eq{eq:cx} (or \eq{eq:cp}) has the Poincare invariance. 
Thus the three-tensor \eq{eq:cx} (or \eq{eq:cp}) can be considered to represent a $D$-dimensional fuzzy infinite flat space \cite{Sasai:2006ua}.

\section{Renormalization procedure}
\label{sec:renormalization}
The philosophy of the Kadanoff-Wilson renormalization procedures is in iterations of coarse graining processes. 
In a discrete system, such a process is generally a discrete step of 
defining new renormalized dynamical variables by averaging over contributions of underlying dynamical variables on nearby sites.
On a fixed lattice, the meaning of nearby sites can obviously be defined by giving a renormalization scale, 
but in a gravitational system, one has to take into account the fact that 
a scale is a dynamical quantity determined by dynamical variables.
Hence, in analogy with a gravitational system,  
a renormalization step in the tensor models should be expressed solely by the dynamical variable, 
and it will generally take a form,
\bea
\label{eq:ren}
C^{R}_{abc}=R\left(C\right)_{abc},
\eea
where $R(C)_{abc}$ is a totally symmetric three-tensor as a function of $C$, 
and $C^R$ is a new renormalized dynamical variable obtained after one step of a renormalization procedure.
The $O(N)$ symmetry \eq{eq:onsym} is assumed to be respected in \eq{eq:ren}. 

In fact, there exist various possibilities of a renormalization procedure $R(C)$, but I assume that
the qualitative behavior in the infrared limit will not depend 
on the details of the choice of $R(C)$, if it is taken appropriately. 
The simplest choice of $R(C)$ \cite{Sasakura:2007ud} is as follows.
Let me consider a configuration of $C$ which is not far from the Gaussian form \eq{eq:cx}. 
Then, if $x,y$ are nearby points,  $C_{x y z} f_z$ gives a function averaged over within the distance of order 
$1/\sqrt{\beta}$ from $x,y$. Similarly, one can consider another averaged function $C_{x' y' z'} f_{z'}$, and 
take the product of these averaged functions,
$C_{x y z} f_z\,C_{x' y' z'} f_{z'}=C_{xyz}C_{x'y'z'}C_{zz'w}f_w$, which will determine a renormalized structure constant. Contracting
the unwanted indices $y$ and $y'$, one obtains a renormalized three-tensor as
\be
\label{eq:renproc}
R(C)_{abc}=C_{ade}C_{bdf}C_{cef}.
\ee 
Note that this expression respects the $O(N)$ symmetry.

It is obvious that there exist various possibilities to insert $C_{abc}$ into \eq{eq:renproc} to change the detailed way of
averaging over. I assume that these changes do not alter the essential properties of 
the infrared behavior of a renormalization procedure. I will show a numerical evidence for
this robustness in Section \ref{sec:example}. In the following section, I will discuss the simplest renormalization 
procedure \eq{eq:renproc}.

In general, the renormalization procedure \eq{eq:ren} does not change the number of degrees of freedom, and hence this is not the same as the 
usual Kadanoff-Wilson renormalization procedure. But as will be discussed in Section \ref{sec:scalartensor}, after a number of 
iterations of the renormalization procedure, the degrees of freedom can be accumulated into a scalar and 
a two-tensor field.

\section{Proof of convergence to Gaussian form}
\label{sec:proof}
Let me first discuss the uniqueness of the Gaussian configurations \eq{eq:cp} up to the second order of momenta
in the exponential, under the assumption of the momentum conservation.
Since $C_{p_1\,p_2\,p_3}$ must be totally symmetric,
the most general expression up to the second order is given by
\bead
\label{eq:cpgen}
C_{p_1\,p_2\,p_3}=&A \exp [
h^\mu p_{1\,\mu}+h^\mu p_{2\,\mu}+h^\mu p_{3\,\mu} \\
&\ \ \ \ \ +
h^{\mu;\nu} p_{1\,\mu} p_{2\,\nu}+h^{\mu;\nu} p_{2\,\mu} p_{3\,\nu}
+h^{\mu;\nu} p_{3\,\mu} p_{1\,\nu}\\
&\ \ \ \ \ +
h^{\mu\nu} p_{1\,\mu} p_{1\,\nu}+h^{\mu\nu} p_{2\,\mu} p_{2\,\nu}+h^{\mu\nu} p_{3\,\mu} p_{3\,\nu}
] \\
&\times \delta^D(p_1+p_2+p_3),
\eead
where 
$h^{\mu;\nu}$ and $h^{\mu\nu}$ are symmetric, and the momentum conservation is expressed by 
the delta function.
Because of $p_1+p_2+p_3=0$, the sum in the first line vanishes.
Moreover, by using $h^{\mu;\nu} (p_1+p_2+p_3)_\mu (p_1+p_2+p_3)_\nu=0$, the 
expression in the second line can be converted to that in the third line. 
Thus \eq{eq:cp} is the most general expression up to 
 the second order of momenta in the exponential.

Now let me apply the renormalization procedure \eq{eq:renproc} to the Gaussian configuration \eq{eq:cp}. 
Noting that the contraction of indices must be taken
for pairs of $p$ and $-p$ in the momentum basis, one obtains 
\bead
R\left(C^g\right)_{p_1\,p_2\,p_3}&=C^g_{p_1 \, q_1\, q_2} C^g_{p_2\, -q_1\, q_3} C^g_{p_3\, -q_2\, -q_3} \\
&=A^3 \delta^D (p_1+p_2+p_3) 
\int d^Dq 
\exp \left(-\alpha(p_1^2+p_2^2+p_3^2+2q^2+2(p_1+q)^2+2(p_2-q)^2)\right)\\
&=A^3 \left[\frac{\pi^D g}{(6 \alpha)^D}\right]^\frac{1}{2} \exp\left(
-\frac{5}{3} \alpha (p_1^2+p_2^2+p_3^2)\right) \delta^D(p_1+p_2+p_3),
\eead
where $g=\hbox{Det}(g_{\mu\nu})$.\footnote{Note that $g$ is for $g_{\mu\nu}$, while $p^2=g^{\mu\nu}p_\mu p_\nu$.}
This expression is again a Gaussian configuration, and the parameters $A,\alpha$ are renormalized as
\bead
\label{eq:renaal}
A_R&=A^3 \left[\frac{\pi^Dg}{(6 \alpha)^D}\right]^\frac{1}{2},\\
\alpha_R&=\frac{5}{3} \alpha.
\eead
This proves that the Gaussian configurations \eq{eq:cx} or \eq{eq:cp} are on the trajectories of 
the renormalization procedure \eq{eq:renproc}. 

In the following, I will show that the Gaussian configurations are the attractors of the three-tensor under the renormalization procedure. 
For simplicity, let me assume that $g_{\mu\nu}=g^{\mu\nu}=\delta_{\mu\nu}$. This simplification 
does not ruin the generality of the discussions below,
because a linear transformation of the momenta in \eq{eq:cp} can transform a general positive-definite 
$g^{\mu\nu}$ to $\delta_{\mu\nu}$.
Let me consider a small deviation from the Gaussian configurations \eq{eq:cp} in the form,
\be
\label{eq:devc}
C_{p_1 \,p_2\,p_3}=A\left(1+f\left(\sqrt{\alpha}p_1,\sqrt{\alpha}p_2,\sqrt{\alpha}p_3\right)\right)
\exp\left( -\alpha(p_1^2+p_2^2+p_3^2)\right)\delta^D(p_1+p_2+p_3),
\ee
where the function $f$ represents a deviation from \eq{eq:cp}, and must be symmetric with respect to its arguments.
Note that this deviation does not violate the translational invariance or the momentum conservation.
The violating case will be discussed in Section \ref{sec:scalartensor}.

Now let me discuss the renormalization of the deviation $f$. 
Let me assume that the deviation is so small 
that the renormalization procedure can be approximated by the linear approximation
in $f$.
The linear approximation is enough to see whether the Gaussian configurations are attractive or not. 
Applying the renormalization procedure \eq{eq:renproc} to \eq{eq:devc}, one obtains
\bead
\label{eq:rendev}
R(C)_{p_1p_2p_3}&=C_{p_1 \, q_1\, q_2} C_{p_2\, -q_1\, q_3} C_{p_3\, -q_2\, -q_3} \\
&\simeq A^3 \delta^D (p_1+p_2+p_3) 
\int d^Dq \left[1+f\left(\sqrt{\alpha} p_1,\sqrt{\alpha} q,\sqrt{\alpha} (-q-p_1)\right)\right.\\
&\left.\ \ \ +f\left(\sqrt{\alpha} p_2,-\sqrt{\alpha} q,\sqrt{\alpha} (q-p_2)\right)
+f\left(\sqrt{\alpha} p_3,\sqrt{\alpha} (p_2-q),\sqrt{\alpha} (q+p_1)\right)\right]\\
&\ \ \ \times \exp \left(-\alpha(p_1^2+p_2^2+p_3^2+2q^2+2(p_1+q)^2+2(p_2-q)^2)\right)\\
&=A^3 \exp\left(
-\frac{5}{3} \alpha (p_1^2+p_2^2+p_3^2)\right) \delta^D(p_1+p_2+p_3) \int d^Dq \exp(-6 \alpha q^2) \\
&\ \ \ \times \left[1+f\left(\sqrt{\alpha}p_1,\sqrt{\alpha}(q+(p_2-p_1)/3),\sqrt{\alpha}(-q+(p_3-p_1)/3)\right)\right.\\
&\ \ \ \ \ \ \ \ +f\left(\sqrt{\alpha}p_2,\sqrt{\alpha}(q+(p_3-p_2)/3),\sqrt{\alpha}(-q+(p_1-p_2)/3)\right)\\
&\left.\ \ \ \ \ \ \ \ +f\left(\sqrt{\alpha}p_3,\sqrt{\alpha}(q+(p_1-p_3)/3),\sqrt{\alpha}(-q+(p_2-p_3)/3)\right)\right], 
\eead
where only terms linear in $f$ are taken from the first to the second 
equalities. After replacing $A,\alpha$ with the renormalized
$A_R,\alpha_R$ in \eq{eq:renaal}, the renormalized deviation $f_R$ can be read from \eq{eq:rendev} as
\bead
\label{eq:fr}
f_R&\left(\sqrt{\alpha_R} p_1,\sqrt{\alpha_R}p_2,\sqrt{\alpha_R}p_3\right)=
\left(\frac{18\alpha_R}{5\pi}\right)^\frac{D}{2}  \int d^Dq\,  \exp\left(- \frac{18 \alpha_R}{5} q^2\right) \\
&\times \Bigg[f\left(\sqrt{\frac{3\alpha_R}{5}}p_1,\sqrt{\frac{3\alpha_R}{5}}(q+(p_2-p_1)/3),\sqrt{\frac{3\alpha_R}{5}}
(-q+(p_3-p_1)/3)\right)\\
&\ \ \ \ \ +(\hbox{cyclic permutations of }p_1,p_2,p_3)\Bigg]. 
\eead

In this paper, I consider an $f$ which contains 
only a finite number of polynomials of $p_i$.
In the coordinate basis, such a deviation corresponds to adding a finite number of polynomials of 
$x_i-x_j$ to the constant $B$ in \eq{eq:cx}. Therefore such a deviation is local in the sense that
it does not change the exponential damping behavior of $C_{x_1x_2x_3}$ at large mutual distances among $x_i$. 
A general deviation with such damping behavior may have an expression with an infinite number of polynomials.
But, because of the damping behavior, it will be enough to consider only the central region, and 
it will be possible to approximate a deviation with
good accuracy by a finite number of polynomials. Then the proof below will be applicable also for such general cases.

Because of the momentum conservation $p_1+p_2+p_3=0$, the expression of $f$ 
in terms of $p_i$ is not unique,  that causes difficulties in the proof. 
This ambiguity can be avoided by expressing $f$ solely in terms of two of the momenta. Since $f$ is symmetric with respect to 
the momentum arguments, there exist three equivalent expressions, each of which can be obtained from each choice of two momenta, 
as
\bead
\label{eq:fbyh}
&f(\sqrt{\alpha}p_1,\sqrt{\alpha}p_2,\sqrt{\alpha}p_3)\Big|_{p_1+p_2+p_3=0}\\
&=f(\sqrt{\alpha}p_1,\sqrt{\alpha}p_2,\sqrt{\alpha}(-p_1-p_2))=\sum_{m,n=0 \atop m+n\neq 1}^{m+n\leq M} \alpha^\frac{m+n}{2}
 h^{\mu_1\ldots \mu_m;\nu_1\ldots \nu_n} p_{1\, \mu_1}
\cdots p_{1\, \mu_m} p_{2\, \nu_1} \cdots p_{2\, \nu_n}\\
&=f(\sqrt{\alpha}(-p_2-p_3),\sqrt{\alpha}p_2,\sqrt{\alpha}p_3)
=\sum_{m,n=0 \atop m+n\neq 1}^{m+n\leq M} \alpha^\frac{m+n}{2} h^{\mu_1\ldots \mu_m;\nu_1\ldots \nu_n} p_{2\, \mu_1}
\cdots p_{2\, \mu_m} p_{3\, \nu_1} \cdots p_{3\, \nu_n}\\
&= \cdots,
\eead
where $M$ is the maximal degree of the polynomials of $p_i$, and $h$ are the coefficients.
These $h$ are symmetric with respect to the indices:
$h^{\mu_1\ldots \mu_m;\nu_1\ldots \nu_n}=h^{\nu_1\ldots \nu_n;\mu_1\ldots \mu_m}$ and
$h^{\mu_{\sigma(1)}\ldots \mu_{\sigma(m)};\nu_{\sigma'(1)}\ldots \nu_{\sigma'(n)}}=
h^{\mu_1\ldots \mu_m;\nu_1\ldots \nu_n}$, where $\sigma,\sigma'$ denote permutations.
Here the right-hand sides of \eq{eq:fbyh} 
do not contain any linear terms with $m+n=1$, because of the momentum conservation, $p_1+p_2+p_3=0$.

Substituting the polynomial form \eq{eq:fbyh} into \eq{eq:fr},
the integration over $q$ of each term will generate terms of lower or equal degrees.
Therefore the renormalization procedure closes with the form in \eq{eq:fbyh},
and it can be expressed as a linear transformation
with an upper triangular matrix,
\be
\label{eq:renlin}
\left(
\begin{matrix}
h_{R\,0}\\
h_{R\,2}\\
\vdots \\
h_{R\,M}
\end{matrix}
\right)
=
\left(
\begin{matrix}
R_{00} & R_{02} & \ldots  & R_{0M} \\
0 & R_{22} & \ldots &R_{2M} \\
0&0& \ddots&\vdots \\
0&0&0 & R_{MM}
\end{matrix}
\right)
\left(
\begin{matrix}
h_0\\
h_2\\
\vdots \\
h_M
\end{matrix}
\right),
\ee
where $h_i$ and $h_{R\,i}$ symbolically represent the coefficients of the terms of degree $i$ in $f$ as in \eq{eq:fbyh}
and $f_R$, respectively,  and $R_{ij}$ symbolically represents
the sub-matrix relating $h_j$ to $h_{R\,i}$. 
The eigenvalues of the linear transformation \eq{eq:renlin} 
determine whether the Gaussian configurations are attractive or
not under the renormalization procedure.
Because of the triangular form of the linear transformation, the eigenvalues are determined solely by the eigenvalues of 
the sub-matrices $R_{ii}$ on the diagonal. 

The matrices $R_{ii}$ on the diagonal represent the contributions from $f$ to $f_R$
which does not change the degrees of the polynomials. 
These contributions can be evaluated simply by ignoring $q$ in the arguments of $f$ in \eq{eq:fr},
and is explicitly given by    
\bead
\label{eq:frequal}
&f^{equal}_R\left(\sqrt{\alpha_R} p_1,\sqrt{\alpha_R}p_2,\sqrt{\alpha_R}p_3\right)\\
&\ \ =
f\left(\sqrt{\frac{3\alpha_R}{5}}p_1,\sqrt{\frac{3\alpha_R}{5}}\frac{p_2-p_1}{3},\sqrt{\frac{3\alpha_R}{5}}
\frac{p_3-p_1}{3}\right)+(\hbox{cyclic permutations of }p_1,p_2,p_3).
\eead
To discuss the matrix $R_{KK}\ (K\leq M)$, 
let me assume that $f$ (and $f^{equal}_R$) contain only polynomials of degree $K$. 
Then \eq{eq:frequal} becomes
\bead
\label{eq:frk}
f^{equal}_R\left(p_1,p_2,p_3\right)=\left(\frac{3}{5}\right)^\frac{K}{2}
f\left(p_1,\frac{p_2-p_1}{3},\frac{p_3-p_1}{3}\right)+(\hbox{cyclic permutations of }p_1,p_2,p_3).
\eead

To discuss the qualitative behavior of the process \eq{eq:frk}, 
let me introduce a norm of $f$ defined by 
\be
\parallel f \parallel \equiv
\underset{\underset{\scriptstyle p_1+p_2+p_3=0}{(p_1)^2,(p_2)^2,(p_3)^2\leq 1}}{\hbox{Max}}\left| 
f(p_1,p_2,p_3)\right|,
\ee
where $|\ |$ denotes an absolute value,
and Max represents taking the maximum value under the conditions below it.
From \eq{eq:frk}, by using 
$| f_1+f_2 | \leq | f_1 | + | f_2 | $,  
one obtains
\bead
\label{eq:ineqk}
\parallel f^{equal}_R \parallel &\leq 3 \left(\frac{3}{5}\right)^\frac{K}{2}
\underset{\underset{\scriptstyle p_1+p_2+p_3=0}{(p_1)^2,(p_2)^2,(p_3)^2\leq 1}}{\hbox{Max}}
\left|f\left(p_1,\frac{p_2-p_1}{3},\frac{p_3-p_1}{3}\right)\right|\\
&\leq  3 \left(\frac{3}{5}\right)^\frac{K}{2} \parallel f \parallel,
\eead
since the three momenta in the arguments of $f$ in the first line of \eq{eq:ineqk} satisfy 
\bead
p_1+\frac{p_2-p_1}{3}+\frac{p_3-p_1}{3}&=0, \\
\left(p_1\right)^2,\left(\frac{p_2-p_1}{3}\right)^2,\left(\frac{p_3-p_1}{3}\right)^2&\leq 1, 
\eead
under $(p_i)^2\leq 1$ and $p_1+p_2+p_3=0$. The inequality \eq{eq:ineqk} proves that, for $K\geq 5$, 
the deviation asymptotically vanishes by iterations of the renormalization procedure.
This implies that the eigenvalues of $R_{KK}\ (K\geq 5)$ are smaller than 1. 

For $K=0,2,3,4$, the inequality \eq{eq:ineqk} is useless because the numerical factor is larger than one.
In these cases, one can explicitly write down all the possible forms of $f$, and explicitly apply \eq{eq:frk} to
obtain the eigenvalues of the matrices $R_{KK}$. For $K=0$, the only form of $f$ is a constant and 
the eigenvalue is obviously 3.
For $K=2$, there exists only one kind of form, $h^{\mu\nu}(p_{1\,\mu}p_{1\,\nu}+p_{2\,\mu}p_{2\,\nu}+p_{3\,\mu}p_{3\,\nu})$,
and the eigenvalues turn out to be $1$. 
For $K=3$, a detailed analysis shows that there exists only one kind of form, and the eigenvalues are
$(3/5)^{3/2}$.
For $K=4$, there exist three kinds of form, and the eigenvalues are $11/25$ and $3/25$. 
The details of these computations
are given in Appendix \ref{app:details}. 

The above discussions show that the eigenvalues of $R_{KK}$ are less than one for $K\geq 3$, but not for $K=0,2$.
This implies that, by iterations of the renormalization procedure, the fluctuations of order $K\geq 3$ 
asymptotically vanishes, while those with $K=0,2$ do not. However, these remaining terms have the forms of a constant and 
$h^{\mu\nu} (p_{1\,\mu}p_{1 \,\nu}+p_{2\,\mu}p_{2 \,\nu}+p_{3\,\mu}p_{3 \,\nu})$, respectively,
which can be absorbed into the redefinition of $A$ and $g^{\mu\nu}$
of the Gaussian configuration \eq{eq:cp}, so that they are deleted from fluctuations.
This proves that the Gaussian configurations \eq{eq:cx} or \eq{eq:cp} are attractive under the renormalization procedure \eq{eq:renproc}.

\section{Numerical examples}
\label{sec:example}
In this section, I will show two numerical examples of how configurations converge to the Gaussian ones.\footnote{The Mathematica program 
used in this section can be downloaded from http://www2.yukawa.kyoto-u.ac.jp/$\sim$sasakura/codes/renormalization.nb.}
To test the renormalization procedure in concrete cases,
the starting configurations are taken from the numerical solutions to the equation of motion derived from the action \eq{eq:action}.
The equation of motion is given by
\be
\label{eq:eom}
-g_0\, C_{abc} + \frac{g_1}{3} 
\left( C_{ade}C_{def}C_{fbc} +\hbox{(cyclic permutations of $a,b,c$)}\right)
-g_2\, C_{ade}C_{bdf}C_{cef}=0.
\ee 
This equation has
a number of numerical solutions corresponding to various dimensional fuzzy tori and spheres \cite{Sasakura:2006pq}.
In this section, as the simplest example, 
I will consider a solution corresponding to a one-dimensional fuzzy torus (a ring), 
which has a one-dimensional translational symmetry.
\begin{figure}
\begin{center}
\includegraphics[scale=.5]{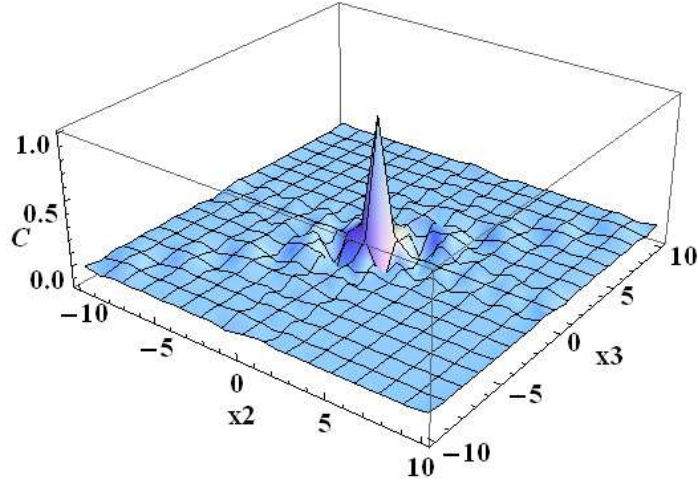}
\caption{The solution $C_{0\,x_2\,x_3}$ representing a fuzzy ring for $L=10$ is plotted in the coordinate basis. The values are normalized by $C_{0\,0\,0}$. 
The coordinate $x$ dual to $p$ is defined by
the discrete Fourier transformation $\sum_{p=-L}^{L} \exp(2 \pi i \, p\, x/(2L+1))$.
}  
\label{fig:cx}
\end{center}
\end{figure}

\begin{figure}
\begin{center}
\includegraphics[scale=.8]{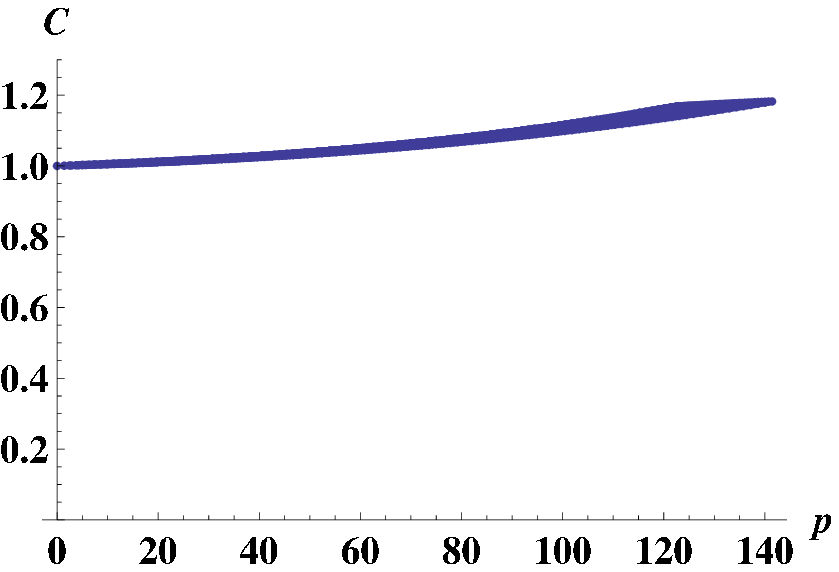}
\hfil
\includegraphics[scale=.8]{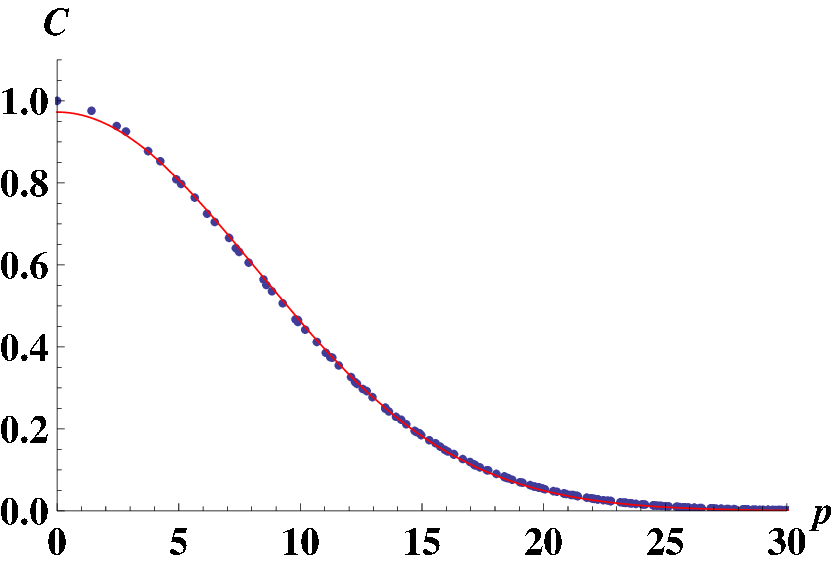}
\caption{The left figure is the plot of the starting configuration which is a solution of \eq{eq:eom} for $g_0=1,g_1=2,g_2=1,L=100$.
The right figure is the plot of the configuration obtained after 10 iterations of the renormalization procedure \eq{eq:renproc}.
The vertical and horizontal axes are $C_{p_1\,p_2\,p_3}$ in the momentum basis 
and $p=\sqrt{p_1^2+p_2^2+p_3^2}$, respectively. The values of $C_{p_1\,p_2\,p_3}$
are normalized by $C_{0\,0\,0}$. The solid line in the right figure is a fit with 
a Gaussian function, $c_0 \exp(- c_1 p^2)$. }
\label{fig:c3}
\end{center}
\end{figure}

Because of the translational symmetry, it is convenient to describe the solutions in the momentum 
basis\footnote{At first sight it is not evident whether the index $p$ below has 
the physical meaning of momentum,
since a space is dynamically generated and there is no preferred frame because of the $O(N)$ 
symmetry \eq{eq:onsym}. In fact,
physical meaning of an index can only be understood by checking the internal consistency of an interpretation 
of a solution as a whole. 
For example, one can see that, in the basis $x$ dual to $p$,
the numerical solution respects locality, i.e. $C_{x_1\,x_2\,x_3}$ is negligible 
unless $x_i$ are nearby locations
as in Figure \ref{fig:cx}. 
This fact ensures that the dual basis $x$ can appropriately be interpreted as a coordinate with a small fuzziness.}. 
To consider a ring, the momentum index is assumed to take only integer values.
A cut-off $L$ is introduced so that the range of the momentum index is finite, $p=-L,-L+1,\ldots,L$.
The one-dimensional translational symmetry can be realized by the momentum conservation
of the solution, i.e. $C_{p_1\,p_2\,p_3}\neq 0$ only if $p_1+p_2+p_3=0$.   
In the left figure of Figure \ref{fig:c3}, a solution for $g_0=1,g_1=2,g_2=1,L=100$ is plotted. 
The form of the solution is very different from
a Gaussian function. Nonetheless, after 10 
iterations of the renormalization procedure, the configuration approaches to a Gaussian configuration as shown in the 
right figure of Figure \ref{fig:c3}.

\begin{figure}
\begin{center}
\includegraphics[scale=.8]{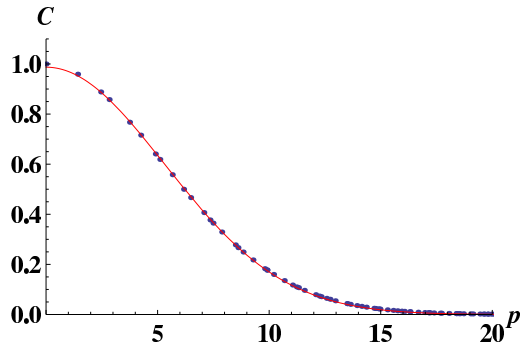}
\caption{The configuration obtained after 10 iterations of the other renormalization process \eq{eq:renanother}.
The solid line is a Gaussian fit.}
\label{fig:another}
\end{center}
\end{figure}

It will be interesting to check the robustness of the convergence to the Gaussian configurations 
under the changes of the renormalization procedure. 
As an example, let me consider another renormalization procedure defined by
\be
\label{eq:renanother}
\tilde R(C)_{abc}=C_{ade}C_{bdf}C_{ceg}C_{fhi}C_{hig}+\hbox{(cyclic permutations of $a,b,c$)},
\ee
which can be obtained by inserting some $C$ to \eq{eq:renproc}.
Since \eq{eq:renanother} is a sum of non-symmetric terms, this renormalization procedure does not seem to have the Gaussian configurations
as stable configurations. Nonetheless, after 10 times of iteration, the configuration  
converges very well to a Gaussian function as shown in Figure \ref{fig:another}. This fact suggests that
the arguments of this paper might be more widely applicable than to the specific renormalization
procedure \eq{eq:renproc}.
Presently, I have no proof of convergence for such general cases, but this character of convergence might be 
related to the so-called central limiting theorem in statistics.

\section{Effective field theory}
\label{sec:scalartensor}
The discussions so far in the previous sections have assumed the translational invariance of the fluctuations.
In this section, the fluctuations are allowed to have a very small momentum $(p_1+p_2+p_3)^2 \ll 1/\alpha$, 
and are assumed to respect the locality, i.e. $C_{x_1\,x_2\,x_3}$ is negligible unless $x_i$ are mutually 
located within the distance of order  
$1/\sqrt{\beta}$.\footnote{The two conditions are mutually compatible. In the coordinate
basis, the first condition implies that fluctuations of $C_{x\,x+a\,x+b}$ have 
wave lengths of order $1/\sqrt{(p_1+p_2+p_3)^2}\gg \sqrt{\alpha}$ as functions of $x$, while
the second one implies that $C_{x\,x+a\,x+b}$ is negligible unless $a^2,b^2 \lesssim 1/\beta \sim \alpha$.}

Let me take the coordinate basis, and assume 
that the whole space is covered by patches of finite sizes well larger than the scale of fuzzyness, 
$1/\sqrt{\beta}\sim \sqrt{\alpha}$,
but well smaller than the wavelength, $1/\sqrt{(p_1+p_2+p_3)^2}\gg \sqrt{\alpha}$.
Then because of the assumption of the locality of $C$, the renormalization procedure \eq{eq:renproc} is
practically a local process within each patch.
Moreover, $C$ can be considered to have the translational invariance with good accuracy in each patch.
Thus the proof in Section \ref{sec:proof} will be applicable to each patch,
and a number of iterations of the renormalization procedure \eq{eq:renproc} on $C$ will result in 
a Gaussian configuration on each patch. Then the collection of the Gaussian configurations of all the patches
will define a Gaussian configuration depending on locations in the whole space.
Such a generalized Gaussian configuration may be expressed as 
\be
\label{eq:gaussgen}
C_{x_1\,x_2\,x_3}=e^{\phi(x_1)+\phi(x_2)+\phi(x_3)} g(x_1)^\frac14 g(x_2)^\frac14 g(x_3)^\frac14
\exp \left(
-\beta \left(d(x_1,x_2)^2+d(x_2,x_3)^2+d(x_3,x_1)^2\right)\right),
\ee
where $\phi(x)$ is a scalar field, $g(x)=\hbox{Det}(g_{\mu\nu}(x))$, and 
$d(x_i,x_j)$ is the geodesic distance between $x_i$ and $x_j$ on a background of a metric 
tensor field $g_{\mu\nu}(x)$.
The above expression \eq{eq:gaussgen} is covariant under the general coordinate transformation in the 
sense that the contraction of indices, $C_{x_1 x_2 x}C_{x x_3 x_4}=
\int d^D x \sqrt{g(x)} e^{2\phi(x)}\cdots$, is invariant, provided that $\phi(x)$ and $g_{\mu\nu}(x)$ transform
covariantly as a scalar and a two-tensor field, respectively. 
The main motivation for the specific form \eq{eq:gaussgen} is that this invariance is necessary 
to identify a part of the $O(N)$ symmetry of the tensor model
with a fuzzy analogue of the diffeomorphism symmetry. 
Note that the covariant expression \eq{eq:gaussgen} is not unique:
$g_{\mu\nu}(x)$ may be rescaled by using $\phi(x)$, e.g. $g_{\mu\nu(x)}\rightarrow e^{\gamma \phi(x)} g_{\mu\nu(x)}$,
and there may be corrections 
in higher orders of $1/\beta$, e.g. $g_{\mu\nu}(x)\rightarrow g_{\mu\nu}(x)+R_{\mu\nu}(x)/\beta$, etc.

In the renormalization procedure \eq{eq:renproc}, the number of degrees of freedom do not change, that is out of
the philosophy of the Kadanoff-Wilson renormalization procedure. However,
the above discussions show that if the above assumptions are met,
the configurations can be well 
approximated by a scalar and a two-tensor field after a number of iterations of the renormalization procedure. 
Then the infrared dynamics of the tensor models will be described by 
an effective field theory of a scalar and a two-tensor field, where
the latter field will be identified with the metric tensor field of the general relativity
because of the fuzzy analogue of the diffeomorphism symmetry.
Indeed, the form \eq{eq:gaussgen} is a generalization of the one presented in the previous papers \cite{Sasakura:2007ud,Sasakura:2008pe}, 
which contains only $g_{\mu\nu}(x)$ and has been used in the numerical analysis of the tensor models
which are fine-tuned to have the Gaussian configurations as classical solutions.
The comparison between the light modes in the tensor models and the general relativity has shown 
some remarkable agreements \cite{Sasakura:2007sv,Sasakura:2007ud,Sasakura:2008pe}. 

On the other hand, the discussions in this section show the possibility that 
a scalar field also appears in the infrared dynamics of the tensor models.
Thus, in general situations, the infrared dynamics of the tensor models will be described in terms of 
scalar-tensor theories of gravity.  
Then it becomes a question why a scalar field has not shown up in the previous
analysis of the fine-tuned tensor models.
This is probably because, while 
the fuzzy analogue of the diffeomorphism invariance coming from the 
symmetry of the tensor model constrains the two-tensor field to appear as a gauge field,
the dynamics of the scalar field strongly depends on the details of the tensor models, and
its appearance as a light mode is not always guaranteed.

\section{Summary, discussions and future prospects}
\label{sec:discussions}
In this paper, I have discussed a renormalization procedure for the tensor models which have 
a totally symmetric real three-tensor as their only dynamical variable.
It has been proven that configurations of certain Gaussian forms are the attractors of the three-tensor
under the renormalization procedure.
Then since these Gaussian forms can be parameterized by a scalar and a two-tensor,
it is argued that, in general situations,
the infrared dynamics of the tensor models will be described by scalar-tensor theories of gravity.

An important assumption in the argument is the locality of a background classical solution and the fluctuations
around it. This is not an evident assumption in the tensor models, because the three-tensor contains
non-local elements. The locality of a background solution can directly be checked by explicitly 
evaluating the three-tensor. On the other hand, the fluctuations around it are generally  
not local. Then what is really assumed is that the non-local fluctuations are not important in the 
infrared dynamics of the tensor models. This is a problem of dynamics, which could be studied 
in numerical manners.

In this paper, the dynamical variable of the tensor models is a totally symmetric real three-tensor. A natural extension
of this is to consider a three-tensor satisfying the generalized hermiticity condition, which was 
considered in the original proposals of the tensor models  
\cite{Ambjorn:1990ge,Sasakura:1990fs,Godfrey:1990dt}. This turns out to be equivalent to consider 
two real three-tensors, one of which is totally symmetric and the other is anti-symmetric.
Thus this extension will inevitably introduce an anti-symmetric two-tensor field by 
similar discussions as in this paper. Then the whole content of fields will agree with
the massless fields in the bosonic string theory \cite{Green:1987sp}. 
It would be highly interesting to study the properties of this new field to see whether this agreement 
is just an accident or has more meaning.

Another interesting direction is to apply the renormalization procedure to the actual analysis 
of the tensor models. 
As explained at the end of the previous section, it is
presently not clear why a scalar field has not shown up in the previous analysis of the tensor models
which are fine-tuned to have Gaussian classical solutions \cite{Sasakura:2007sv,Sasakura:2007ud,Sasakura:2008pe}.
Probably this is because the scalar field can appear as a light field only on limited conditions.
In this regard, the tensor model with the action \eq{eq:action} will be interesting to study, because 
it is known to have various physically interesting classical backgrounds
without fine-tuning of the coupling constants \cite{Sasakura:2006pq}.
The model will also be interesting as an example to study the characteristics of the scalar-tensor theories describing the tensor models.
The forms of scalar-tensor theories of gravity depend on the physical motivations \cite{Brans:1996tp}. 
Among them, the idea of Brans-Dicke \cite{Brans:1961sx} seems very interesting in the present context, 
because, in the scenario of emergent spaces and gravity, the gravitational constant will
be a dynamical quantity which will be determined not only by a background spacetime but also by matter distributions. 


\appendix
\section{The eigenvalues of $R_{KK}$ for $K=2,3,4$}
\label{app:details}
In this appendix, for $K=2,3,4$, I will explicitly obtain all the independent forms of the functions $f(p_1,p_2,p_3)$, 
which are symmetric with respect to $p_i$ and are defined only on the momentum conserved plane, $p_1+p_2+p_3=0$.
Then $f_R^{equal}$ is computed by \eq{eq:frk} to obtain the eigenvalues of $R_{KK}$.

The tensor $h$ in this appendix is assumed to have the symmetries appropriate for \eq{eq:fbyh},
i.e. $h^{\mu_1\ldots \mu_m;\nu_1\ldots \nu_n}=h^{\nu_1\ldots \nu_n;\mu_1\ldots \mu_m}$ and
$h^{\mu_{\sigma(1)}\ldots \mu_{\sigma(m)};\nu_{\sigma'(1)}\ldots \nu_{\sigma'(n)}}=
h^{\mu_1\ldots \mu_m;\nu_1\ldots \nu_n}$, where $\sigma,\sigma'$ denote permutations.

\subsection{$K=2$}
Apparently, for $K=2$, there exist two functions which are symmetric with respect to $p_i$ as
\bead
\label{eq:k2app}
f_1(p_1,p_2,p_3)&=h_1^{\mu\nu} \left(p_{1\,\mu}p_{1 \,\nu}+p_{2\,\mu}p_{2 \,\nu}+p_{3\,\mu}p_{3 \,\nu}\right),\\
f_2(p_1,p_2,p_3)&=h_2^{\mu;\nu} \left(p_{1\,\mu}p_{2 \,\nu}+p_{2\,\mu}p_{3 \,\nu}+p_{3\,\mu}p_{1 \,\nu}\right).
\eead 
However these are not independent, 
because, by substituting $p_3=-p_1-p_2$ into \eq{eq:k2app}, these functions turn out to have the same form as
\bead
f_1(p_1,p_2,-p_1-p_2)&=2 h_1^{\mu\nu} \left(p_{1\,\mu}p_{1 \,\nu}+p_{2\,\mu}p_{2 \,\nu}+p_{1\,\mu}p_{2 \,\nu}\right),\\ 
f_2(p_1,p_2,-p_1-p_2)&=-h_2^{\mu;\nu} \left(p_{1\,\mu}p_{1 \,\nu}+p_{2\,\mu}p_{2 \,\nu}+p_{1\,\mu}p_{2 \,\nu}\right).
\eead
Thus there exist only one independent form.
Substituting $f_1$ into \eq{eq:frk}, one finds
\be
f_{1\,R}^{equal}(p_1,p_2,-p_1-p_2)=\left(\frac{3}{5}\right)^\frac{K}{2} \frac{5}{3}\, f_1(p_1,p_2,-p_1-p_2)=f_1(p_1,p_2,-p_1-p_2).
\ee
Thus the eigenvalues of $R_{22}$ are 1.

\subsection{$K=3$}
For notational simplicity, let me first introduce some short-hand notations,
\bead
p^3&\equiv h^{\mu\nu\rho} p_\mu p_\nu p_\rho,\\
(p_1^2, p_2)&\equiv h^{\mu\nu;\rho} p_{1\,\mu}p_{1\,\nu}p_{2\,\rho},
\eead
and so on.
These are convenient, because one can compute without taking so much care of the tensors $h$. 
The following formulas hold as usual:
\bead
(p_1+p_2)^3&=p_1^3+3p_1^2 p_2+3p_1 p_2^2+p_2^3, \\
\left((p_1+p_2)^2,p_3\right)&=(p_1^2+2p_1p_2+p_2^2,p_3)=(p_1^2,p_3)+2 (p_1p_2,p_3)+(p_2^2,p_3),
\eead
where $p_1^2p_2\equiv h^{\mu\nu\rho} p_{1\,\mu}p_{1\,\nu}p_{2\,\rho}$, and so on.
With these notations, there exist two apparently independent forms,
\bea
\label{eq:fk31}
f_1&=&p_1^3+p_2^3+p_3^3=-3 (p_1^2p_2+p_1p_2^2), \\
f_2&=&(p_1^2,p_2)+\hbox{(permutations of $p_1,p_2,p_3$)}=(p_1^2,p_2)+2(p_1p_2,p_1+p_2)+(p_2^2,p_1),
\label{eq:fk32}
\eea
where I have used $p_3=-p_1-p_2$.
In fact, $f_1$ and $f_2$ are not independent, because, by substituting $h^{\mu\nu;\rho}=h^{\mu\nu\rho}$ into \eq{eq:fk32}, one
can derive the same form as \eq{eq:fk31}.
Thus it is enough to consider only $f_2$, and, by substituting it into \eq{eq:frk}, one obtains 
\bea
f_{2\,R}^{equal}=\left(\frac{3}{5}\right)^\frac{K}{2} f_2.
\eea
Thus the eigenvalues of $R_{33}$ are $(3/5)^{3/2}$.

\subsection{$K=4$}
With similar simplified notations as above, there exist three apparently different forms,
\bead
f_1&=p_1^4+p_2^4+p_3^4=2(p_1^4+2p_1^3p_2 +3 p_1^2 p_2^2+2p_1p_2^3+p_2^4),\\
f_2&=(p_1^3,p_2)+\hbox{(permutations of $p_1,p_2,p_3$)}\\
&=-2(p_1^3,p_1)-2(p_2^3,p_2)-(p_1^3,p_2)-(p_2^3,p_1)-3(p_1p_2(p_1+p_2),p_1+p_2),\\
f_3&=(p_1^2,p_2^2)+\hbox{(cyclic permutations of $p_1,p_2,p_3$)}\\
&=(p_1^2,p_1^2)+(p_2^2,p_2^2)+3(p_1^2,p_2^2)+2(p_1p_2,p_1^2+p_2^2),
\eead
where I have used $(p_1^2,p_2^2)=(p_2^2,p_1^2)$, which comes from $h^{\mu\nu;\rho\sigma}=h^{\rho\sigma;\mu\nu}$.
In fact, $f_1$ can be obtained from $f_2$ by substituting $h^{\mu\nu\rho;\sigma}=h^{\mu\nu\rho\sigma}$, and 
therefore is not independent.
Another substitution of  
$h^{\mu\nu\rho;\sigma}=h^{\mu\nu;\rho\sigma}+h^{\rho\mu;\nu\sigma}+h^{\nu\rho;\mu\sigma}$ 
into $f_2$ results in a new form,
\be
f_4=(p_1^2,p_1^2)+(p_2^2,p_2^2)+(p_1^2,p_2^2)+2(p_1p_2,p_1p_2)+2(p_1p_2,p_1^2+p_2^2).
\ee

Of course, this $f_4$ is not an independent form because this is derived from $f_2$. But as will be shown below, this form is needed to
express $f_{3\,R}^{equal}$. For this reason, it is convenient to divide the whole space of the form $f_2$ into the form of $f_4$ and the rest, and 
regard $f_4$ as an independent form.  
By subtracting $f_3$ from $f_4$, one obtains a simpler expression,
\be
f_5=(p_1^2,p_2^2)-(p_1p_2,p_1p_2).
\ee 
Thus there exist three independent forms, $\tilde f_2,f_3,f_5$, where $\tilde f_2$ denotes the rest of $f_2$ explained above.

By substituting these forms into \eq{eq:frk}, one obtains
\bead
f_{2\,R}^{equal}&=\left(\frac{3}{5}\right)^\frac{K}{2} \frac{11}{9} f_2=\frac{11}{25} f_2,\\
f_{3\,R}^{equal}&=\left(\frac{3}{5}\right)^\frac{K}{2} \left( \frac{11}{9} f_3-\frac{16}{9} f_5 \right)=\frac{11}{25}f_3-\frac{16}{25}f_5,\\
f_{5\,R}^{equal}&=\left(\frac{3}{5}\right)^\frac{K}{2} \frac{1}{3} f_5=\frac{3}{25} f_5.
\eead 
Thus the eigenvalues of $R_{KK}$ are $11/25$ and $3/25$.

\end{document}